\documentclass[12pt]{article}
\usepackage[legalpaper,margin=2.1cm]{geometry}
\usepackage{amssymb}
\usepackage{graphicx}

\title{Heavy baryon spectroscopy from lattice QCD}
\author{M. Padmanath \thanks{Electronic address : Padmanath.M@physik.uni-regensburg.de} \\
	Instit\"ut  f\"ur  Theoretische Physik, Universit\"at Regensburg, \\ D-93040 Regensburg, Germany 
	}

\date{}
\begin{document}

\maketitle

\begin{abstract}
In this report, the most recent and precise estimates of masses of ground state baryons
using lattice QCD are discussed. Considering the prospects in the heavy baryon sector, lattice 
estimates for these are emphasized. The first and only existing lattice determination of the 
highly excited $\Omega_c$ excitations in relation to the recent LHCb discovery is also discussed. 
\end{abstract}

\section{Introduction}

Since its inception, heavy hadron physics continues to be in the limelight of scientific 
interests in understanding the nature of strong interactions. While heavy mesons have 
been studied extensively both experimentally and theoretically~\cite{Brambilla:2014jmp,Olsen:2015zcy,Prelovsek:2016dmi}, 
studies on heavy baryons remained dormant. In this respect, the year 2017 featured two 
important landmarks in the heavy baryon physics. First of this is the unambiguous observation 
by LHCb collaboration of five new narrow $\Omega_c$ resonances in $\Xi^{+}_{c}K^{-}$ 
invariant mass distribution in the energy range between $3000-3120$ MeV~\cite{Aaij:2017nav}. 
Four out of these five resonances were later confirmed by Belle collaboration 
\cite{Yelton:2017qxg}. The second landmark is the discovery of a doubly charmed baryon, 
$\Xi^{+}_{cc} (ccu)$ with a mass of $3621.40\pm0.78$ MeV by LHCb Collaboration 
\cite{Aaij:2017ueg}. Anticipating the discovery of many more hadrons (including baryons) 
from the huge data being collected at LHCb and Belle II, heavy hadron spectroscopy 
using {\it ab-initio} first principles methodology such as lattice QCD is of great importance. 

Lattice QCD has been proven to be a novel non-perturbative technology in investigating the 
physics of low energy regime of QCD. Remarkable progress has been achieved over past ten years
in making large volume simulations with physical quark masses, impressive statistical precision 
and good control over the systematic uncertainties \cite{Durr:2008zz,Namekawa:2013vu,Borsanyi:2014jba,Alexandrou:2017xwd}. 
In this report, a collection of lattice determinations of baryon masses that are well below allowed strong 
decay thresholds are summarized. A recent and only existing calculation of excited $\Omega_c$ baryons 
is discussed and qualitative comparison with the experiment is made. 

\section{Lattice methodology}

Hadron spectroscopy on the lattice proceeds through evaluation of Euclidean two point correlation 
functions, 
\begin{equation}
C_{ij}(t_f-t_i)=\langle O_{i}(t_f)O^{\dagger}_{j}(t_i) \rangle=\sum_{n=1} \frac{Z_i^{n}Z_j^{n*}}{2 E_n}e^{-E_n(t_f-t_i)}~,
  \label{eq:2-1}
\end{equation}
between different hadronic currents ($O_{i}(t)$) that are carefully built to respect the quantum 
numbers of interest. A generic baryon current or interpolator has a structure 
\begin{equation}
O_{i}({\bf x},t)=\epsilon_{abc} S^{\alpha\beta\delta}_i({\bf x}) q_{1,\alpha}^{a}({\bf x}) q_{2,\beta}^{b}({\bf x}) q_{3,\delta}^{c}({\bf x}),
  \label{eq:2-2}
\end{equation}
where $q_j$ are the quark fields, $\epsilon$ is the color space anti-symmetrizing Levi-Civita tensor 
and $S$ carries all the flavor and spatial structure of the interpolator that determines the quantum 
information. $C_{ij}(t_f-t_i)$ are evaluated on lattice QCD ensembles that are generated via Monte 
Carlo techniques. A general practice is to compute matrices of correlation functions between a basis 
of carefully constructed interpolating currents $O_{i}(t)$ and solving the generalized eigenvalue problem (GEVP)~\cite{Michael:1985ne,Luscher:1985dn,Blossier:2009kd}
\begin{equation}
C_{ij}(t)v_j^n(t-t_0)=\lambda^n(t-t_0)C_{ij}(t_0)v_j^n(t-t_0).
  \label{eq:2-3}
\end{equation}
Hadron energies ($E_n$) are extracted from non-linear fits to the large time behavior of the eigenvalues
$\lambda^n(t-t_0)$. The eigenvectors ($v_j^n(t-t_0)$) are related to the operator state overlaps 
($Z_i^{n} = \langle O_{i}|n\rangle$) that carry the quantum information of the propagating state. 
Basic principles remain the same as above, while details of the methodology differ between different 
groups in the lattice community. e.g. lattice ensembles being used in the study, lattice formulation of 
action for the fermion and the gauge fields, the hadron interpolators, different degree of control over 
the lattice systematics, etc. The success of lattice investigations is reflected in mutual agreement 
of the results they provide and their agreement with experiments. 

All results presented in this report are estimated within the single hadron approximation, where only 
three quark interpolators (as in eqn. (\ref{eq:2-2})) are considered in the analysis and neglects effects 
of any nearby strong decay thresholds. This is a justifying assumption for most of the baryons discussed 
in this report, considering the fact that all of them are deeply below the respective lowest strong decay 
thresholds. Results for those baryons, which might be influenced by any nearby threshold effects will be 
alerted in the respective discussions.

\begin{figure}[h]
\begin{center}
\includegraphics[scale=0.65]{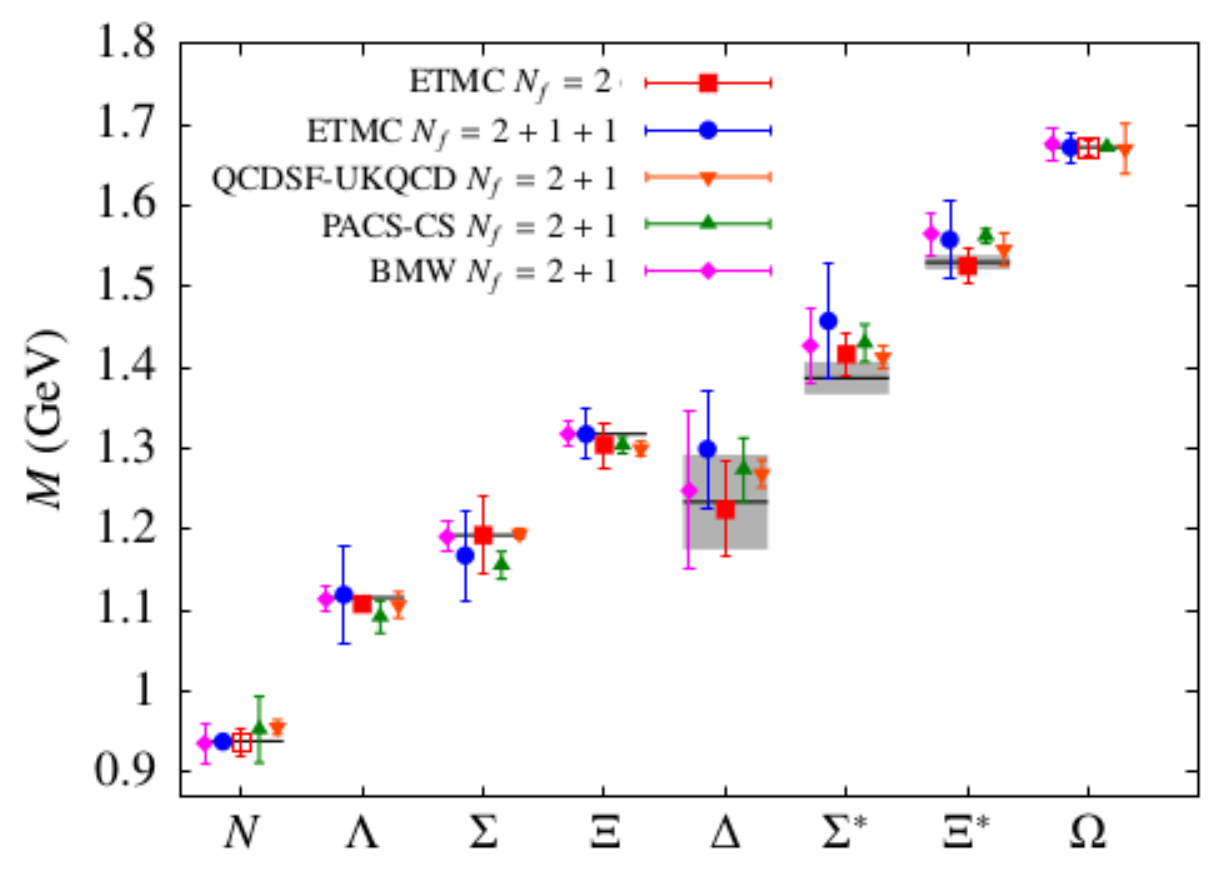}
\includegraphics[scale=0.55]{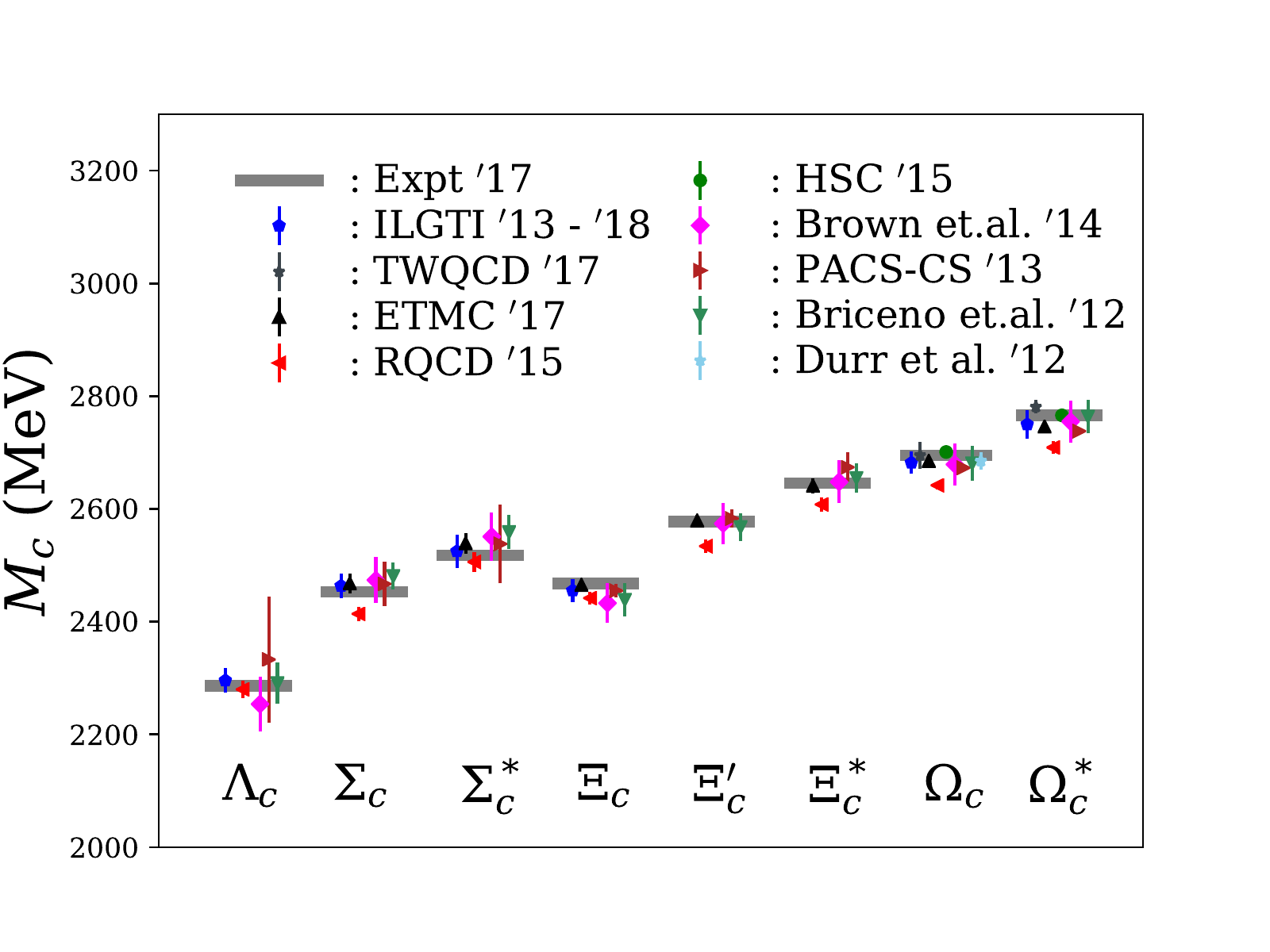}
\end{center}
\caption{Left: (adapted from Ref. \cite{Alexandrou:2017xwd}) summary of lattice estimates for positive parity 
light and strange baryons from selected lattice investigations - ETMC $N_f$=2 \cite{Alexandrou:2017xwd}, 
ETMC $N_f$=2+1+1 \cite{Alexandrou:2014sha}, QCDSF-UKQCD $N_f$=2+1 \cite{Bietenholz:2011qq}, PACS-CS 
$N_f$=2+1 \cite{Namekawa:2013vu} and BMW $N_f$=2+1 \cite{Durr:2008zz}. Right: summary of lattice estimates 
for positive parity singly charm baryons - ILGTI '13-'18 \cite{Basak:2012py,Mathur:2018rwu}, TWQCD '17 
\cite{Chen:2017kxr}, ETMC '17 \cite{Alexandrou:2017xwd}, RQCD '15 \cite{Bali:2015lka}, HSC '15 
\cite{Padmanath:2015bra,Padmanath:2015jea}, Brown {\it et al} '14 \cite{Brown:2014ena}, PACS-CS '13 
\cite{Namekawa:2013vu}, Brice\~no {\it et al} '12 \cite{Briceno:2012wt}, D\"urr {\it et al} '12 
\cite{Durr:2012dw}.}\label{fig:3-1}
\end{figure}

\section{Results}

{\em\it Light, strange and singly charm baryons}:
We begin our discussion with some benchmark calculations of baryon ground states that are experimentally 
well determined. In Fig. \ref{fig:3-1}, a summary of lattice QCD estimates for the positive parity 
light baryon ground states (figure adapted from Ref. \cite{Alexandrou:2017xwd}) are presented on 
the left and for positive parity singly charm baryon ground states are shown on the right. Most of 
the baryons being discussed are deeply bound and stable to strong decays. Their masses as determined 
from the discrete energy spectrum on the lattice agree quite well with experiments. Agreement 
between all the lattice estimates with varying degree of control over the systematics involved in 
respective calculations and with the experiments demonstrate the power of lattice QCD techniques in 
making reliable predictions. However, lattice estimates for masses of baryon resonances, such as 
$\Delta$, $\Sigma^*$ and $\Xi^*$ that can decay strongly, are less rigorous. They demand a computation 
of correlation matrices build out of baryon interpolators (as in eqn. \ref{eq:2-2}) plus baryon-meson 
interpolators (corresponding to the allowed strong decay modes). The masses of baryon resonances then 
have to be inferred from the infinite volume scattering matrices build from the discrete spectrum 
extracted from such correlation matrices. Such investigations are being practiced extensively by many 
collaborations to understand various mesonic resonances (see Ref. \cite{Prelovsek:2016dmi}), while 
existing lattice investigations of baryon resonances in this direction are limited to a few 
\cite{Lang:2016hnn, Andersen:2017una}.
 
\begin{figure}[h]
\begin{center}
\includegraphics*[width=0.49\textwidth,clip]{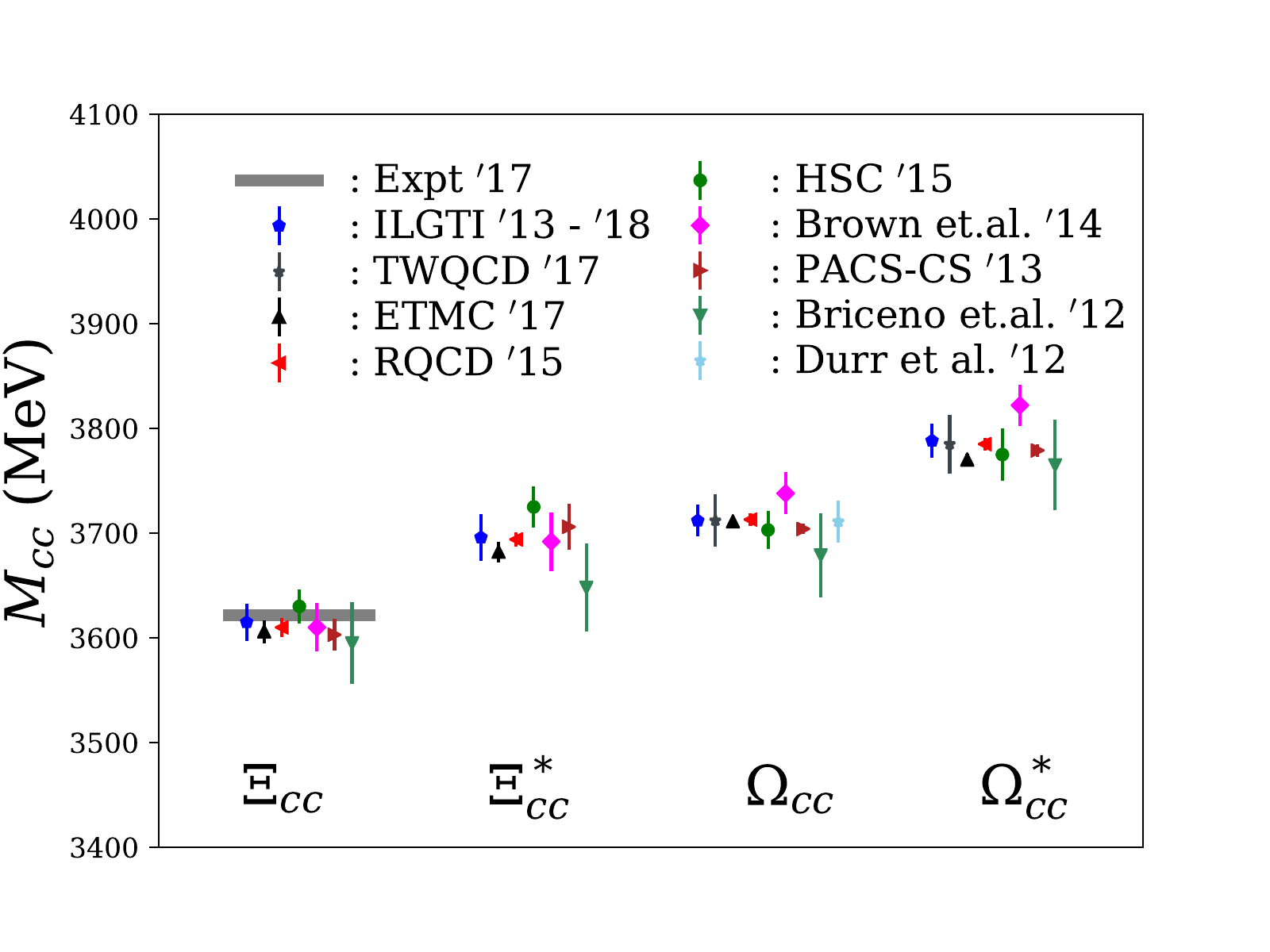}
\includegraphics*[width=0.49\textwidth,clip]{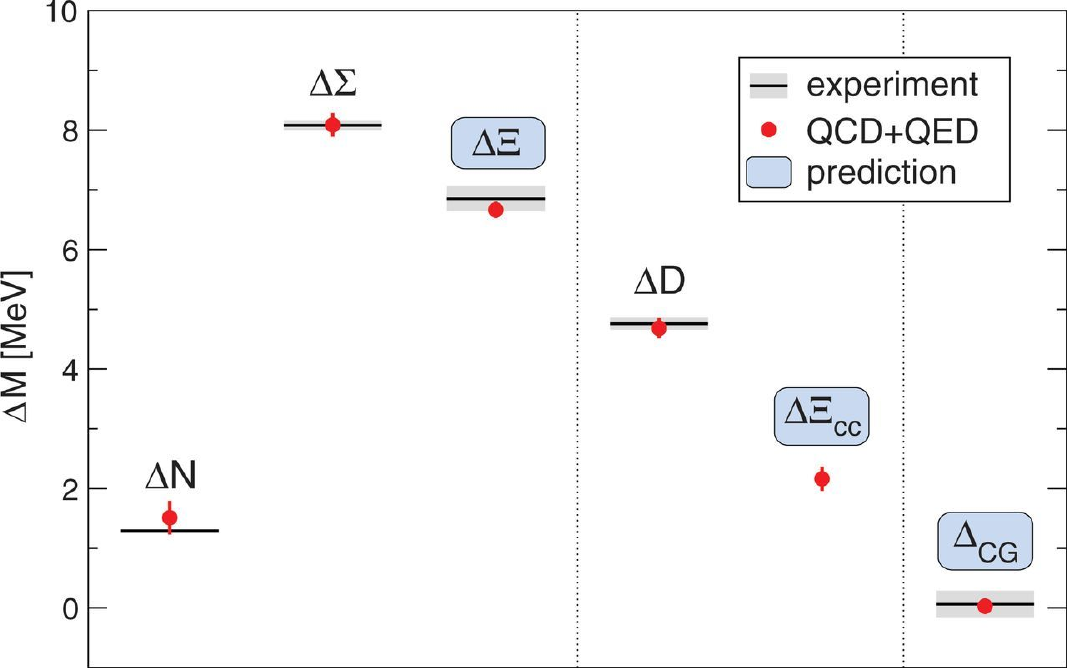}
\end{center}
\caption{Left: summary of lattice estimates for positive parity doubly charm baryons. References as 
given in Fig. \ref{fig:3-1} caption. Right: Hadron isospin splittings as determined by BMW collaboration 
\cite{Borsanyi:2014jba}.}\label{fig:3-2}
\end{figure}

{\em\it Doubly heavy baryons}:
In Fig. \ref{fig:3-2}, a summary of lattice QCD estimates for positive parity doubly charm baryon ground 
states on the left is presented. For the $\Xi_{cc}$($1/2^+$) baryon, good agreement between all lattice estimates 
(all of which predates the LHCb-discovery \cite{Aaij:2017ueg}) and with LHCb estimate is quite evident from the figure. 
At this point, the reader is reminded of the observation of another baryon resonance by SELEX collaboration 
in 2002 \cite{Mattson:2002vu} at a mass of 3519(1) MeV, which is addressed as a $\Xi_{cc}$($1/2^+$) baryon. All 
lattice estimates, being well above this energy, disfavors this observation. The right figure shows a summary 
of baryon isospin splittings as calculated by BMW collaboration \cite{Borsanyi:2014jba}. This 
calculation involved lattice QCD and QED computations with four non-degenerate fermion flavors to 
estimate the isospin mass splitting in the nucleon, $\Sigma$, $\Xi$, $D$ and $\Xi_{cc}$ isospin 
multiplets. Precise estimation of the neutron-proton isospin splitting and the other known splittings 
demonstrate the reliability of these estimates. In this calculation, the isospin splitting of 
$\Xi_{cc}$($1/2^+$) baryon was estimated to be 2.16(11)(17) MeV. This excludes the possibility that LHCb 
and SELEX candidates for $\Xi_{cc}$($1/2^+$) baryon are isospin partners. 

\begin{figure}[h]
\begin{center}
\includegraphics[width=0.5\textwidth,clip]{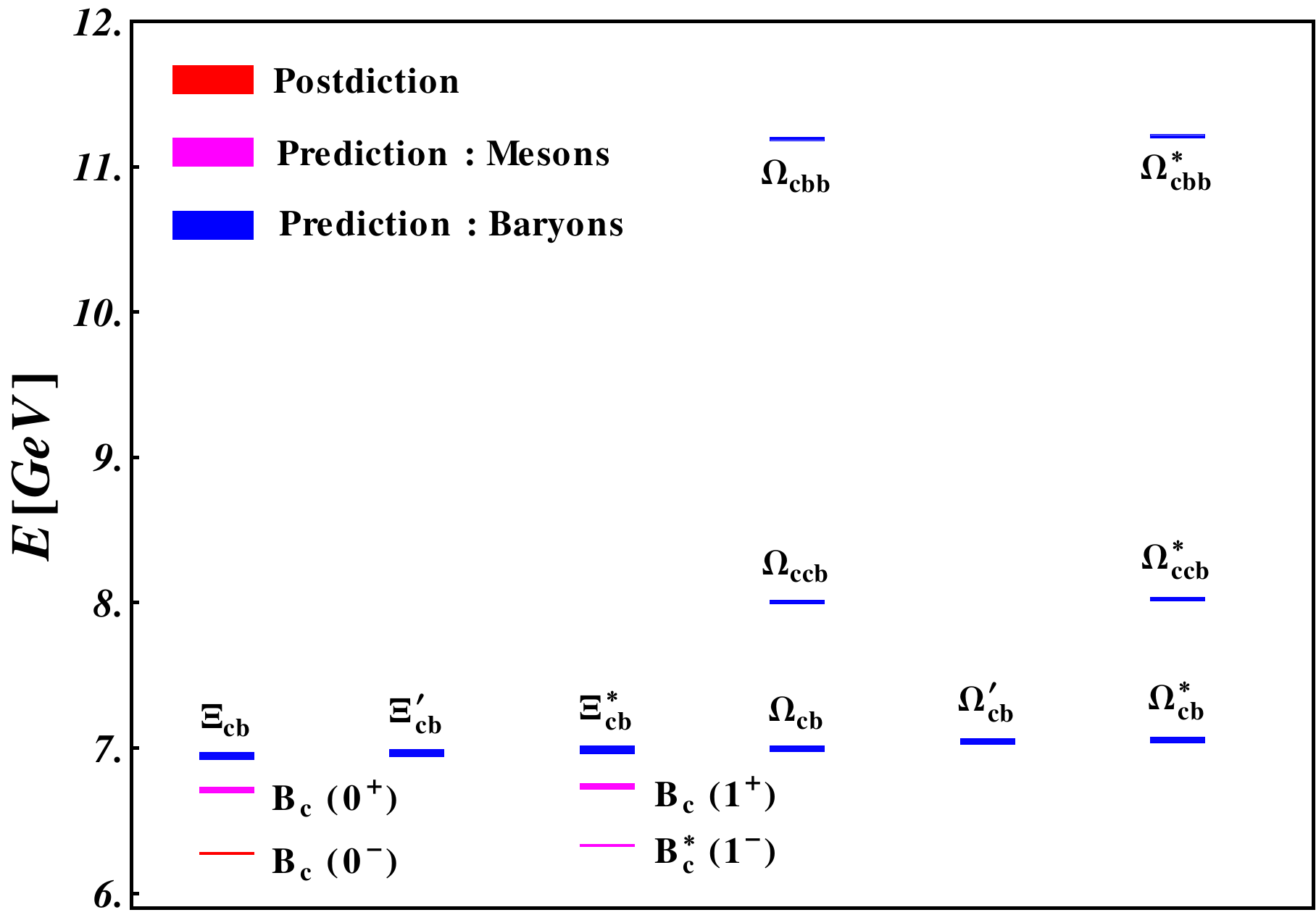}
\includegraphics[width=0.49\textwidth,clip]{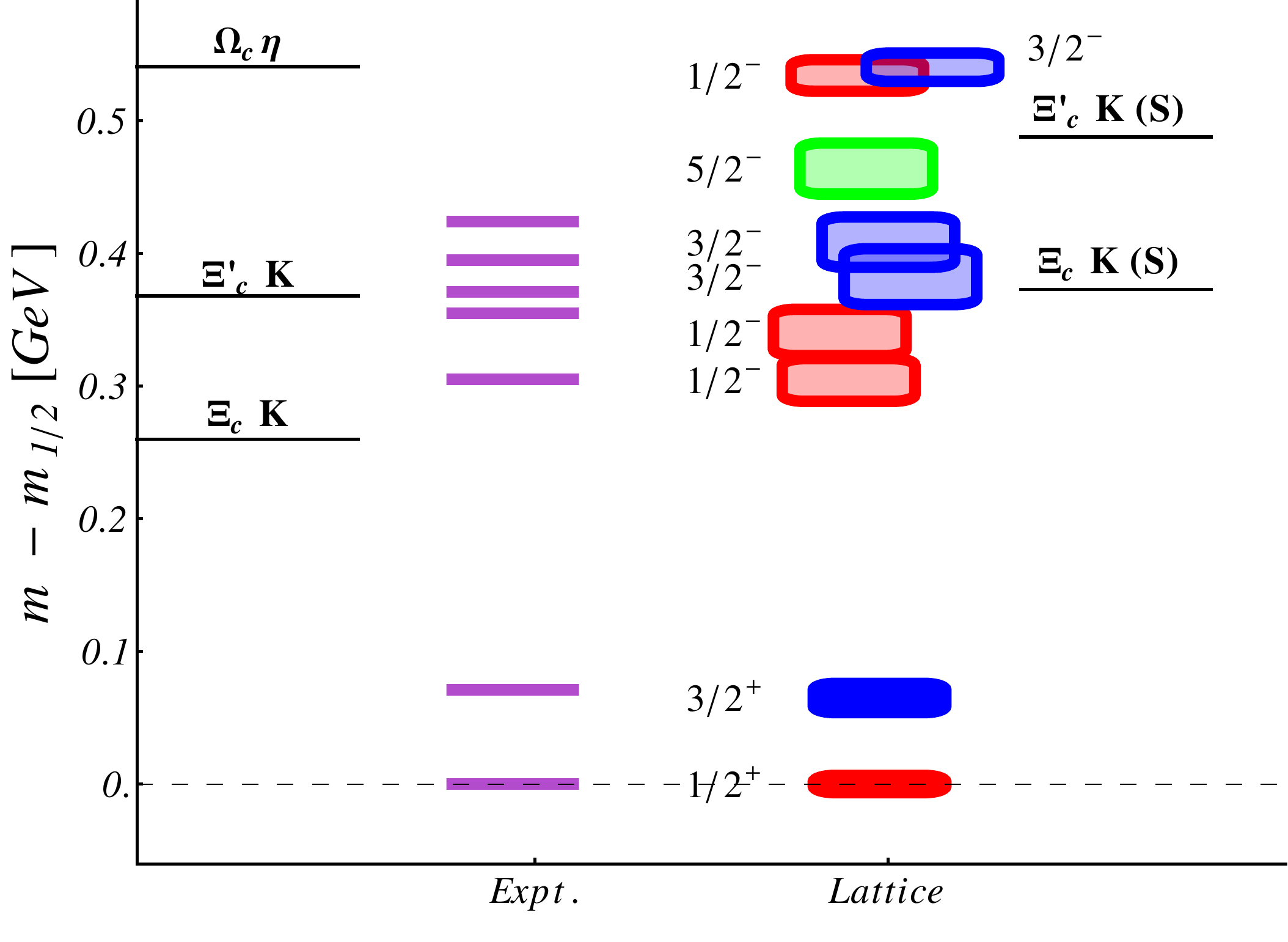}
\end{center}
\caption{Left: summary of lattice estimates for low lying charmed-bottom hadrons as determined 
in Ref. \cite{Mathur:2018epb}. Right: Comparison plot from Ref. \cite{Padmanath:2017lng} between the lattice estimates and 
the experimental values for the energies of $\Omega_c$ excitations.}\label{fig:3-3}
\end{figure}

Estimates for other doubly charm baryons, that are yet to be discovered, can also be observed to 
be very well determined and consistent between different lattice calculations from the left of Fig. \ref{fig:3-2}. 
Anticipating a near future discovery of the charmed-bottom hadrons at LHCb, on the left of Fig. \ref{fig:3-3} 
lattice predictions for such hadrons from a recent investigation \cite{Mathur:2018epb} are shown. 
The lattice prediction for only know charmed-bottom hadron, $B_c$ meson, is found to be in agreement 
with the experiment, while the lattice predictions for other channels considered are consistent 
with another preceding calculation \cite{Brown:2014ena} with less control over systematics. 

{\em\it Excited baryons}: As discussed in the introduction, one of the major landmark in the year 2017 
is the LHCb discovery of five narrow $\Omega_c$ resonances in $\Xi^{+}_{c}K^{-}$ invariant mass 
distribution in the energy range between $3000-3120$ MeV~\cite{Aaij:2017nav}. Following this discovery, 
Belle collaboration has confirmed four out of these five excited states \cite{Yelton:2017qxg}. Many more highly excited baryons 
are coming into light with more discoveries. e.g. the observation of a $\Omega^{*-}$($3/2^-$) candidate 
with a mass of 2012.4(9) MeV by Belle collaboration \cite{Yelton:2018mag}, which is in very good agreement 
with lattice prediction for such a baryon \cite{Edwards:2012fx,Engel:2013ig}. 
Below we discuss the first and only existing lattice investigation of highly excited $\Omega_c$ resonances 
(Ref. \cite{Padmanath:2017lng}) that predicts the five excited $\Omega_c$ baryons as observed by LHCb.

Following a detailed baryon interpolator construction procedure as invented in Ref. \cite{Basak:2005ir,Edwards:2011jj}
a large basis of baryon interpolators, that is expected to extensively scan the radial as well 
as orbital excitations, are built. By solving the GEVP for correlation matrices constructed out 
of these interpolators on a lattice ensemble with $m_{\pi}\sim391$ MeV (for details see \cite{Padmanath:2015jea}), 
one extract the $\Omega_c$ baryon spectrum on the lattice. The right of Fig. \ref{fig:3-3} shows
a comparison of the lattice energy estimates for the lowest nine $\Omega_c$ excitations 
with the seven experimentally observed $\Omega_c$ resonances. The relevant strong decay thresholds 
in the infinite volume are shown as black lines on the left, whereas the black lines on the right 
indicate the relevant non-interacting levels on the lattice. The lowest two levels represent the 
well known $1/2^+$ and $3/2^+$ excitations. Lattice estimates for these excitations agree well 
with the experiment. In the energy region, where the five narrow resonances were observed, 
lattice predicts exactly five levels. Of these five excitations, four are in good agreement with 
the experiment, while the fifth is possibly a $5/2^-$ baryon related to the remaining higher 
lying experimental candidate. The operator state overlaps $Z^n_i$ (see eqn. (\ref{eq:2-1})) 
indicate these five states to be the $p$-wave excitations \cite{Padmanath:2017lng}. 

Considering the exploratory nature of this first study, investigating $\Omega_c$ baryon spectrum 
on multiple lattice ensembles with close to physical $m_{\pi}$ and larger volumes would be 
an immediate extension. It would also be an interesting direction to extract the infinite volume 
scattering matrices considering the allowed baryon-meson scattering channels in the analysis 
of desired quantum channels in appropriate lattice ensembles. However, the presence of a valence 
heavy quark, the absence of any valence light quarks and the resonance widths being quite 
narrow ($<10$ MeV) \cite{Aaij:2017ueg} indicates our estimates to be robust with such extensive investigations. 

\section{Summary}
Over the past decade, lattice QCD has availed multiple precision determinations of the ground state 
baryon masses using full QCD lattice ensembles with good control over the systematic uncertainties. 
A summary of lattice determinations of various baryons along with their masses from the experiment, 
where available, are given in Figs. \ref{fig:3-1}, \ref{fig:3-2} and \ref{fig:3-3}. The only 
existing exploratory lattice determination of the highly excited $\Omega_c$ states in relation to 
the recent LHCb discovery and its possible extensions are also discussed. 

\section{Acknowledgements}
I would like to thank my collaborators N. Mathur, R. G. Edwards, M. J. Peardon and S. Mondal. I also
express my thanks to the organizers and the participants of the Bled workshop for various interesting 
and insightful discussions. I acknowledge the support from the EU under grant no. MSCA-IF-EF-ST-744659 
(XQCDBaryons) and Deutsche Forschungsgemeinschaft Grant No. SFB/TRR 55.

\end{document}